\documentclass[a4paper,12pt]{article}
\usepackage{amsfonts,amsthm,amsmath,amssymb,graphicx,hyperref,color,youngtab}

\def\res{\mathop{\text{Res}}\limits}

\def\a{\alpha}
\def\b{\beta}

\def\g{\gamma}

\newcommand{\be}{\begin{equation}}
\newcommand{\bea}{\begin{eqnarray}}

\newcommand{\ee}{\end{equation}}
\newcommand{\eea}{\end{eqnarray}}

\newcommand{\nn}{\nonumber}

\def\g{\gamma}

\begin{document}

\makeatletter
\@addtoreset{equation}{section}
\makeatother
\renewcommand{\theequation}{\thesection.\arabic{equation}}
\vspace{1.8truecm}

{\LARGE{ \centerline{\bf Non-Perturbative String Theory from AdS/CFT}  }}  

\vskip.5cm 

\thispagestyle{empty} 
\centerline{ {\large\bf Robert de Mello Koch$^{a,b,}$\footnote{{\tt robert@neo.phys.wits.ac.za}}, Eunice Gandote$^{b,}$\footnote{{\tt eunice@aims.edu.gh}} and
Jia-Hui Huang$^{a,}$\footnote{{\tt huangjh@m.scnu.edu.cn}} }}

\vspace{.4cm}
\centerline{{\it ${}^{a}$ School of Physics and Telecommunication Engineering},}
\centerline{{ \it South China Normal University, Guangzhou 510006, China}}

\vspace{.4cm}
\centerline{{\it ${}^{b}$ National Institute for Theoretical Physics,}}
\centerline{{\it School of Physics and Mandelstam Institute for Theoretical Physics,}}
\centerline{{\it University of the Witwatersrand, Wits, 2050, } }
\centerline{{\it South Africa } }

\vspace{1truecm}

\thispagestyle{empty}

\centerline{\bf ABSTRACT}

\vskip.2cm 
The large $N$ expansion of giant graviton correlators is considered.
Giant gravitons are described using operators with a bare dimension of order $N$.
In this case the usual $1/N$ expansion is not applicable and there are contributions to the correlator that are 
non-perturbative in character.
By writing the (square of the) correlators in terms of the hypergeometric function ${}_2F_1(a,b;c;1)$, we are able to rephrase the $1/N$ expansion of the correlator as a semi-classical expansion for a Schr\"odinger equation.
In this way we are able to argue that the $1/N$ expansion of the correlator is Borel summable and that it exhibits a parametric Stokes phenomenon as the angular momentum of the giant graviton is varied. 

\setcounter{page}{0}
\setcounter{tocdepth}{2}
\newpage
\tableofcontents
\setcounter{footnote}{0}
\linespread{1.1}
\parskip 4pt

{}~
{}~

\section{Introduction}

${\cal N}=4$  super Yang-Mills theory is holographically dual to string theory on asymptotically AdS$_5\times$S$^5$ spacetime\cite{Maldacena:1997re,Witten:1998qj,Gubser:1998bc}.
This implies that all excitations appearing in the spectrum of string theory must appear in the CFT Hilbert space.
The usual perturbative spectrum (which consists of supergravity excitations, as well as closed strings) is captured by the planar limit of the dual CFT\cite{Berenstein:2002jq}.
There are also many non-perturbative objects, including branes\cite{McGreevy:2000cw,Grisaru:2000zn,Hashimoto:2000zp,Balasubramanian:2001nh,Corley:2001zk} and new spacetime geometries\cite{Lin:2004nb,Corley:2001zk,Berenstein:2004kk}, an interesting example being spacetimes containing black holes, that must be found in the CFT Hilbert space\cite{Balasubramanian:2005mg,Balasubramanian:2008da}.
These non-perturbative configurations are dual to operators with a bare dimension that grows parametrically with $N$ ($\sim N$ for branes or $\sim N^2$ for new spacetime geometries).
To explain how this works, consider the half-BPS sector where a useful basis for the operators of the theory is given by the Schur polynomials.
Consider a Schur polynomial labeled by a Young diagram consisting of a single column, $\chi_{(1^J)}(Z)$ of $J$ boxes.
For $J\sim O(1)$ the operator is dual to a collection of (point like) KK-gravitons.
As $J$ is increased to $J\sim O(\sqrt{N})$ long single trace operators dual to stringy states start to participate.
Increasing $J$ further to $O(N)$ we obtain a giant graviton brane.
Thus, the dual to the single CFT operator $\chi_{(1^J)}(Z)$ transitions through different physical descriptions (particles, strings and branes) as the parameter $J$ is varied.
It is natural to ask how these different partial representations are combined into a single coherent description.

The character of the large $N$ expansion changes as we transition between these different partial representations.
For $J\ll\sqrt{N}$ we can take the usual 't Hooft limit and the large $N$ theory is just the planar limit.
As $N$ goes to infinity and $J\gg \sqrt{N}$ one must sum much more than just the planar diagrams (see \cite{Balasubramanian:2001nh,Garner:2014kna} for clear and relevant discussions).
For this reason we will refer to these limits as large $N$ but non-planar limits.
In large $N$ but non-planar limits one does not have the usual $1/N$ expansion.
The ribbon graph expansion is not of much help because enormous combinatorial factors imply that the usual higher genus suppression is overwhelmed by the sheer number of diagrams of a given topology\cite{Balasubramanian:2001nh}.
Different trace structures do mix and it is not at all clear how the large $N$ expansion can usefully be organized.
This is a key question that we wish to address, albeit in the limited setting of a specific example.
A nice class of correlators that we will use to explore this issue are three point functions of ${1\over 2}$-BPS operators as well as extremal $n$-point functions of ${1\over 2}$-BPS operators.
There are rigorous non-renormalization theorems\cite{Baggio:2012rr} that prove that these correlators do not receive 't Hooft coupling corrections. Thus, they can be computed exactly, in the free field theory limit.
Even this problem is one of considerable complexity, due to the very large dimensions of the operators.
Fortunately, using group representation theory, this problem has been solved exactly, as we briefly review in Section \ref{correlators}. 
Our goal here is to explain the structure of the $1/N$ expansion for some correlation functions of giant graviton branes.
In this way we will take a small first step towards defining the structure of the $1/N$ expansion in large $N$ but non-planar limits.

Since $\hbar$ of the dual gravitational system is $1/N$, the different large $N$ limits that can be taken lead to different classical configurations of the gravitational theory. 
This is inline with conventional wisdom: when performing a path integral quantization there are many possible saddle points so that typically a quantum system has many perturbative series, each associated to a different classical configuration. 
These series are the basic building block in many computations. 
Although summing a few terms gives a good approximation, these series are almost always divergent. 
One needs a theory that can organize these different series into a coherent description of the quantum system. 
This is precisely what the theory of resurgence does. 
The first step entails converting the divergent series into meaningful objects by Borel resummation in the perturbation parameter. 
Typically one considers a loop expansion and the small parameter is $\hbar$. 
The second step entails exhibiting a relation between the different series, which is manifested through the Stokes phenomenon\footnote{The Stokes phenomenon is the basic fact that, in general, Borel resummations are discontinuous along rays in the complex plane. These rays are the Stokes and anti-Stokes lines.}. 
This relation implies that, given a specific series, the discontinuities of its Borel transform encode the information about other series in the problem. 
In this way, one can (for example) synthesize the usual perturbative expansion, together with the expansions in the (typically many) different instanton sectors, to recover exact results.
From this point of view, the Stokes lines of the perturbative expansion simply demarcate where contributions from other saddle points become dominant.

Given this discussion, it seems that resurgence has a crucial role to play in understanding the large $N$ limit of Yang-Mills theories. 
Specifically, resurgence should be relevant to understand how the different representations (i.e. the different possible large $N$ limits) fit together to provide a complete and coherent description.
If the ideas of resurgence are relevant, there should be a Stokes phenomenon present as the parameter $J$ (and not $\hbar$) is varied.
As a first step in exploring this possibility, we will look for and exhibit this Stokes phenomenon in this paper. 
To approach this problem we use the exact WKB method\cite{Voros}. 
The different perturbative series that appear are the WKB series around different classical trajectories. The basic objects are the (Borel resummed) perturbative series in $\hbar$. 
The series can be characterized by two types of data: their classical limit and their discontinuity structure\cite{Voros,DP1,DP2}, which is encoded in the action of the so-called Stokes automorphisms. 
A simple characterization of the Stokes discontinuities is in terms of Voros symbols, which are simply the exponent of the WKB series. 
Our analysis starts with the observation that the (square of the) correlators we compute can be expressed in terms of the hypergeometric ${}_2F_1(a,b,c,x)$ function. 
This is a useful observation because the differential equation obeyed by the hypergeometric function is easily mapped into a Schr\"odinger equation, which can be approached using an exact WKB analysis. 
The relevant Schr\"odinger equation has $1/N$ playing the role of $\hbar$ so that the WKB expansion of the wave function of this Schr\"odinger equation gives the $1/N$ expansion of our correlator. 
Fortunately, the relevant Schr\"odinger equation has been studied in detail in \cite{KawaiTakei,AokiandTanda,Tanda,att,Aoki}. In particular, the Voros symbols have been studied and their singularity structure in the WKB plane is well understood.
The relevant WKB solutions have been proved to be Borel summable\cite{Koike}.
The solutions do exhibit Stokes phenomena in the parameter $J$ and this has been studied in detail: the Stokes lines and Stokes regions for this equation can be described quite explicitly and connection formulas relating solutions in different Stokes domains are known. 
This implies that the singularities of the Borel transforms of the WKB solutions are well understood and that the Alien calculus for this problem is completely worked out\cite{KawaiTakei,AokiandTanda,Tanda,att,Aoki}. These are the only ingredients needed to give the trans-series expansion for the hypergeometric function and hence of our giant graviton correlators.
In Section \ref{GGWKB} the exact WKB method is applied to unravel the structure of the large $N$ expansion of the giant graviton correlation functions.

A key result of this article is the expansion of extremal $n$-point correlation functions of normalized Schur polynomial operators $O_J$, labeled either by a single column or a single row containing $J$ boxes with $J$ of order $N$. 
It is useful to introduce the parameters $j\equiv {J\over N}$ which are held fixed as we take $N\to\infty$. The expansion is of the form
\bea
\langle O_{J_1}\cdots O_{J_k}O^\dagger_{J_1+\cdots+J_k}\rangle =
{e^{\alpha N}\over\sqrt{N}}\sum_{n=k-1}^\infty c_n N^{-n}\label{basicresult}
\eea
The coefficients $\alpha$ and $c_n$ are functions of the fixed parameters $j_1,j_2,\cdots j_k$. We find that $\alpha$ can be both positive and negative. 
The series (\ref{basicresult}) is an asymptotic series.
For the special case of three point functions we discuss the Borel summation of the series and gives the Stokes region in which the resummation converges. 
The details of the Stokes regions depend on the parameters (functions of the $j_i$s) appearing in the Schr\"odinger equation, so that we can dissect the (complex) parameter space into regions with the topology of the Stokes graph constant in each region. 
The boundary of the region relevant for the three giant graviton correlator has a transparent physical interpretation and corresponds to points at which a giant shrinks to zero size or expands to maximal size. 
These are exactly the limits of the giant graviton description, so that this parametric Stokes phenomenon does indeed seem to be connected to the transition from one physical representation to another.
Identifying the coupling $g_s=N^{-1}$ we see that (\ref{basicresult}) is a particularly simply transeries with a single nonperturbative term parameter $e^{\alpha N}=e^{\alpha\over g_s}$.
These non-perturbative contributions have been identified\cite{Hirano:2018xmh} with instantons in the tiny graviton matrix model\cite{SheikhJabbari:2004ik} description of giant gravitons.
Finally in Section \ref{discuss} we discuss our results and suggest some possible directions for further study.

\section{CFT Correlators}\label{correlators}

The space of half-BPS representations can be mapped to the space of Schur polynomials of $U(N)$, that is, to the space of Young diagrams characterizing representations of $U(N)$\cite{Corley:2001zk}. 
There are rigorous theorems\cite{Baggio:2012rr} that imply that extremal correlation functions of Schur polynomials do not receive any 't Hooft coupling corrections and hence they are given exactly by their values in free field theory.
The Schur polynomials that correspond to giant graviton branes have a single column with order $N$ boxes, while those corresponding to dual giant graviton branes have a single row with order $N$ boxes\cite{Corley:2001zk}.
Computing correlators of these operators is still a highly non-trivial task, even in the free field theory limit, because the number of fields in each operator is going to infinity as we take $N\to\infty$.
Fortunately, using techniques based on group representation theory, this problem has been completely solved in \cite{Corley:2001zk,Corley:2002mj} for operators constructed using a single field (say $Z$) and in \cite{Bhattacharyya:2008rb} for operators constructed using more than one matrix (see also \cite{Kimura:2007wy,Brown:2007xh,Brown:2008ij}).
We give a quick review of these results in this section and then use them to explore different possible behaviors of these correlators at large $N$. 

Let $V$ denote the $N$ dimensional vector space carrying the fundamental representation of $U(N)$.
The space ${\rm Sym}(V^{\otimes n})$ is also a representation of $U(N)$ but it carries in addition a commuting action of $S_n$.
These actions can be simultaneously diagonalized leading naturally to the operators of interest to us, the Schur polynomials.
After diagonalizing, the representations of both groups can be labeled by a Young diagram that has $n$ boxes.
A further consequence is that the two point function is also diagonalized.
The simplest way to achieve the diagonalization is by using a projection operator. 
The Schur polynomials are given by
\bea
\chi_R(Z)={\rm Tr}(P_R Z^{\otimes n})
\eea
where $P_R$ is a projection operator
\bea
P_R={1\over n!}\sum_{\sigma\in S_n}\chi_R(\sigma )\sigma
\eea
projecting from $V^{\otimes n}$ to $R$.
The two point function is given by the trace of a product of projectors, which can be evaluated exactly to find
\bea
   \langle \chi_R(Z)(x_1)\chi_S(Z^\dagger)(x_2)\rangle = {f_R\delta_{RS}\over |x_1-x_2|^{2n_R}}
\eea
where $R$ has $n_R$ boxes.
The number $f_R$ appearing on the right hand side of this equation is equal to the product of the factors, one for each box in the Young diagram. 
Recall that a box in row $i$ and column $j$ has factor $N-i+j$.

The computation of a three point function of Schur polynomials follows immediately from the above result, upon using the  product rule enjoyed by Schur polynomials
\bea
   \chi_R(Z)\chi_S(Z)=\sum_T g_{RST}\chi_T(Z)
\eea
where $g_{RST}$ is known as a Littlewood-Richardson number.
Consequently the three point function is given by
\bea
   \langle \chi_R(Z)(x_1)\chi_S(Z)(x_2)\chi_T(Z^\dagger )(x_3)\rangle 
={f_T g_{RST}\over |x_1-x_3|^{2n_R}|x_2-x_3|^{2n_S}}
\eea

It is rather natural to study operators with a two point function normalized to 1.
The normalized version of the Schur polynomial is given by
\bea
   O_R(x)={\chi_R(x)\over \sqrt{f_R}}
\eea
The normalized three point correlator is given by
\bea
   \langle O_R(Z)(x_1) O_S(Z)(x_2) O_T(Z^\dagger )(x_3)\rangle 
=\sqrt{f_T\over f_R f_S}{g_{RST}\over |x_1-x_3|^{2n_R}|x_2-x_3|^{2n_S}}
\eea
These are the correlators we will study in this article.
In what follows the spacial dependence plays no role and consequently from now on we omit it.
This dependence is easily reinstated using simple dimensional analysis.
Use $A_J$ to denote the antisymmetric representation with $J$ boxes (i.e. the Young diagram $A_J$ has a single column) and $S_J$ to denote the symmetric representation with $J$ boxes (i.e. $S_J$ has a single row).
It is straight forward to see that
\bea
    \langle O_{A_{J_1}}O_{A_{J_2}}O_{A_J}\rangle = \sqrt{(N-J_1)!(N-J_2)!\over (N-J)! N!}\label{corr}
\eea
where $J=J_1+J_2$.
We stress that this expression is the exact answer, valid for any values of $J_1$ and $J_2$.
For operators in the planar limit we would hold $J_1,J_2$ fixed as we take $N\to\infty$ in which case expanding the correlator leads to a well behaved power series in $N^{-1}$.
To make this point we can consider $J_1=J_2=2$ in which case
\bea
\langle O_{A_{2}}O_{A_{2}}O_{4}\rangle &=& \sqrt{(N-2)(N-3)\over (N-1)N}
=\sqrt{\left( 1-2{1\over N}\right)\left(1-3{1\over N}\right)\over 1-{1\over N}}\cr
&=&1-2 {1\over N}-{1\over N^2}-{1\over N^3}-\frac{3}{2}{1\over N^4}-3 {1\over N^5}-7 {1\over N^6}
+O\left(N^{-7}\right)
\eea
This expansion in $1/N$ converges absolutely in the range $0\le {1\over N}<{1\over 3}$, that is, for $N>3$.
This planar limit is the regime in which we study perturbative string theory, so it is perhaps not too surprising that we can perform a $1/N$ expansion.
This result is however, better than we may have expected: most perturbative expansions are only asymptotic expansions.
If we were to increase $J_1$ and $J_2$ the radius of converges would shrink further.

We could also consider the case that $J_1=O(N)$ with ${J_1\over N}$ fixed and much less than $1$. The result (\ref{corr}) is exact, so it continues to hold in this limit. Since $J_1$ is order $N$, $O_{A_{J_1}}$ is a giant graviton. We can then take $J_2=2n=O(1)$, so that $O_{A_{J_2}}$ is some collection of point gravitons. 
The correlator (\ref{corr}) then describes the emission or absorption of gravitons by a giant graviton.
In this case we find
\bea
\langle O_{A_{J_1}}O_{A_{J_2}}O_{A_J}\rangle \simeq \left(1-{J_1\over N}\right)^n+O(N^{-1})
\eea
and there is again no obstacle to carrying out a $1/N$ expansion. 
A simple case study is provided by taking $J_2=2$ in which case we have
\bea
\langle O_{A_{J_1}}O_{A_{J_2}}O_{A_J}\rangle &=&\sqrt{(N-J_1)!(N-2)!\over (N-J_1-2)! N!}
=\sqrt{(N-J_1)(N-J_1-1)\over N(N-1)}\cr
&=&
(1-j_1)-\frac{j_1}{2 N}+\frac{j_1 (3 j_1-4)}{(8-8 j_1) N^2}+O\left(N^{-3}\right)
\eea
where $j_1={J_1\over N}$.
This is again an absolutely convergent expansion for ${1\over N}<1-j_1$.

To obtain giant graviton correlators we should set $J_i=Nj_i$ and hold $j_i$ fixed as we take $N\to\infty$. Giant gravitons are spherical D3-brane states which are not part of the perturbative string spectrum\cite{McGreevy:2000cw,Grisaru:2000zn,Hashimoto:2000zp}, so we might expect that this correlator does not have a $1/N$ expansion.
In this limit the normalized correlator behaves as\cite{Brown:2006zk}
\bea
\langle O_{A_{J_1}}O_{A_{J_2}}O_{A_J}\rangle = \sqrt{(N-J_1)!(N-J_2)!\over (N-J)! N!}\simeq e^{- N j_1 j_2}\label{NPcorr}
\eea
The exponential on the right hand side of the above correlator does not admit a ${1\over N}$ expansion and it therefore constitutes a genuine non-perturbative contribution.
The corresponding result for the dual giant graviton correlator is
\bea
\langle O_{S_{J_1}}O_{S_{J_2}}O_{S_J}\rangle = \sqrt{(N+J_1-1)!(N+J_2-1)!\over (N+J-1)! (N-1)!}\simeq e^{N j_1 j_2}\label{NPcorr2}
\eea
which is again a non-perturbative contribution.
One of the main results of this article is the trans-series expansion of (the square of) these giant graviton correlators.
The starting point for this analysis uses Gauss' Hypergeometric Theorem, which says
\bea
{}_2F_1(a,b;c;1)={\Gamma (c)\Gamma (c-a-b)\over\Gamma (c-a)\Gamma (c-b)}
\eea
Clearly then, we can write
\bea
\left(\langle O_{A_{J_1}}O_{A_{J_2}}O_{A_J}\rangle\right)^2 =\,\, {}_2F_1(-J_1,J_2;N-J_1+1;1)
\eea
\bea
\left(\langle O_{S_{J_1}}O_{S_{J_2}}O_{S_J}\rangle\right)^2 =\,\,{}_2F_1(J_1,-J_2;N+J_1;1)
\eea
It is possible to transform the hypergeometric differential equation into the Schr\"odinger equation and then use any of the techniques developed for quantum mechanics.
To explore the structure of the $1/N$ expansion of the giant graviton correlators, we will use known results for the exact WKB expansion for the hypergeometric function.

The language of Schur polynomials generalizes to the case of multi matrix models.
The Schur polynomials are replaced by restricted Schur polynomials.
For concreteness focus on restricted Schur polynomials constructed from two complex matrices $Z$ and $Y$.
These restricted Schur polynomials are labeled by three Young diagrams\footnote{In general we also need some extra multiplicity labels. These labels however will not be needed for the giant graviton correlators we study so they will be omitted from our discussion for simplicity.} $\chi_{R,(r,s)}(Z,Y)$.
For an operator constructed using $n$ $Z$ fields and $m$ $Y$ fields, the Young diagram $r$ has $n$ boxes, $s$ has $m$ boxes and $R$ has $n+m$ boxes\cite{Bhattacharyya:2008rb}. A giant graviton operator would be given by the restricted Schur polynomial $\chi_{A_{n+m},(A_n,A_m)}(Z,Y)$, while a dual giant graviton operator is given by $\chi_{S_{n+m},(S_n,S_m)}(Z,Y)$.
The normalized correlator of three giant gravitons is given by\cite{Bhattacharyya:2008xy}
\bea
&&\langle \chi_{A_{n_1+m_1},(A_{n_1},A_{m_1})}\chi_{A_{n_2+m_2},(A_{n_2},A_{m_2})}
\chi^\dagger_{A_{n_{12}+m_{12}},(A_{n_{12}},A_{m_{12}})}\rangle\cr\cr
&&=\sqrt{(N-n_1-m_1)!(N-n_2-m_2)!\over N! (N-n_{12}-m_{12})!}\sqrt{n_{12}!m_{12}!(n_1+m_1)!(n_2+m_2)!\over n_1!n_2!m_1!m_2!(n_{12}+m_{12})!}\label{corr3}
\eea
This result is a product of two square root factors. The first factor has the same form as the one matrix result.
The second factor is always $\le 1$. 
To see this, consider the binomial expansion of 
$$ (1+x)^m = \sum^m_{k=0}\left(^m_k\right)x^k,\qquad {\rm where}\qquad \left(^m_k\right)={m!\over k!(m-k)!}$$
By comparing the coefficient of $x^{r+s}$ coming from the expansion of $(1+x)^m$ times the expansion of $(1+x)^n$ to the coefficient of $x^{r+s}$ coming from the expansion of $(1+x)^{m+n}$, we learn that
$$ \left(^m_r\right)\left(^n_s\right)\quad + \quad {\rm non\, negative\, integers}\quad = \left(^{m+n}_{r+s}\right).$$
Thus
$$ {\left(^m_r\right)\left(^n_s\right)\over\left(^{m+n}_{r+s}\right)} \le 1,$$
which proves that the second factor is $\le 1$. Notice that when $m_1=m_2=0$, the second factor is identically equal to 1 so that our result correctly reduces to the one matrix result we discussed above. 
It is equally easy to compute the correlation function for three dual giant gravitons.
The result is\cite{Bhattacharyya:2008xy}
\begin{eqnarray} 
&&\langle \chi_{S_{n_1+m_1},(S_{n_1},S_{m_1})}\chi_{(S_{n_2+m_2},(S_{n_2},S_{m_2})}
\chi^\dagger_{S_{n_{12}+m_{12}},(S_{n_{12}},S_{m_{12}})}\rangle\cr\cr\cr
&&=\sqrt{(N+n_{12}+m_{12}-1)!(N-1)!\over (N+m_1+n_1-1)! (N+n_{2}+m_{2}-1)!}\sqrt{n_{12}!m_{12}!(n_1+m_1)!(n_2+m_2)!\over n_1!n_2!m_1!m_2!(n_{12}+m_{12})!}\label{corr4}
\end{eqnarray}
It is again easy to verify that if we set $m_1=0=m_2$, we recover the correct one matrix result.

There are again a number of physical processes described by our correlators (\ref{corr3}) and (\ref{corr4}): a three point correlators of point like graviton or of string states, a three point function involving two giant gravitons and one string or point like graviton or a three point correlator involving only giant gravitons.
We will only quote the result for a correlator involving three giant gravitons.
Setting $n_1=n_2=n=N\mathfrak{n}$ and $m_1=m_2=N\mathfrak{m}$ we find
\bea
\langle \chi_{A_{n+m},(A_{n},A_{m})}\chi_{A_{n+m},(A_{n},A_{m})}\chi^\dagger_{A_{2n+2m},(A_{2n},A_{2m})}\rangle\sim e^{-2N(\mathfrak{n}+\mathfrak{m})}\sqrt{2(\mathfrak{n}+\mathfrak{m})\over\pi N\mathfrak{n}\mathfrak{m}}
\eea
\begin{eqnarray} 
\langle \chi_{S_{n+m},(S_{n},S_{m})}\chi_{(S_{n+m},(S_{n},S_{m})}\chi^\dagger_{S_{2n+2m},(S_{2n},S_{2m})}\rangle\sim e^{2N(\mathfrak{n}+\mathfrak{m})}\sqrt{2(\mathfrak{n}+\mathfrak{m})\over\pi N\mathfrak{n}\mathfrak{m}}
\end{eqnarray}
Both of these correlators again are quantities that can not be expanded in a power series in $N^{-1}$.

It is also interesting to consider extremal $n$-point functions since these are also protected. The result is
\bea
   \langle O_{R_1} O_{R_2}\cdots  O_{R_k} O_T^\dagger\rangle 
=g_{R_1 R_2\cdots R_k T} \sqrt{f_T\over f_{R_1} f_{R_2}\cdots f_{R_k}}
\eea
where $g_{R_1 R_2\cdots R_k T}$ counts how many times $T$ appears in the tensor product of $R_1$ with $R_2$ with $\cdots$ with $R_k$.
We will see, in explicit examples considered later, that these correlators also exhibit interesting behavior that is non-perturbative with respect to the $1/N$ expansion.

\section{Review of Exact WKB}\label{EWKB}

In this section we will review the exact WKB solution to the Schr\"odinger equation\cite{Voros}.
This is useful because as we review in the next section, the hypergeometric differential equation can be mapped into the Schr\"odinger equation with a specific potential.
The exact WKB method starts from the usual WKB expansion to write the wave function as an infinite series.
The Borel sums of these WKB solutions exhibit parametric Stokes phenomena\cite{Aoki}, which is a Stokes phenomena in the asymptotic behavior of WKB solutions with a change in the parameters of the problem\footnote{One could also consider Stokes phenomenon arising as a consequence of changing $\hbar$. This is typically considered when analyzing the Borel resummation of WKB solutions and it is not what we are considering here.}. 
The space of parameters can be partitioned into regions by the Stokes graph.
The vertices of the Stokes graphs are singular points as well as turning points associated to the Schr\"odinger equation.
The Borel sum of the WKB solutions converge within each face of the Stokes graph, but are discontinuous across the Stokes lines.
This parametric Stokes phenomena is nicely captured in Voros coefficients, which describe the relative normalization of wave functions normalized\footnote{By a wave function normalized at a point $x_0$, we mean the overall amplitude of the wave function $\psi(x)$ is chosen so that $\psi(x_0)=1$.} at well chosen distinct points. 
The whole analysis can be phrased in terms of Stokes' automorphims and Alien derivatives, introduced by Ecalle\cite{Ecalle} in his theory of resurgence.
This analysis explicates the singularities of the Borel sum and these are the seeds of the non-perturbative contribution to the wave function.

\subsection{Orientation}

Broadly speaking the collection of ideas that we are drawing on go under the name of resurgence.
Since much of the background maybe a little unfamiliar, in this section we will give a very brief overview of the relevant ideas. 
For very helpful background reading, aimed at theoretical physicists, we suggest the reader consults \cite{Marino:2015yie,Dorigoni:2014hea,Aniceto:2018bis,Dunne:2016qix}.

Use $g$ to denote the coupling constant.
The perturbative expansion of an interesting observable ${\cal O}$ will take the form
\bea
{\cal O}=\sum_{n=0}^\infty c_n g^n
\eea
Typically the coefficients $c_n$ grow as $n!$ so that this series does not converge,  but rather it defines an asymtotic expansion.
In this situation, we would typically carry out a Borel resummation.
This is a two step process, in which we first perform a Borel transform of the sum and then we perform a Laplace transform.
The Borel transform of a given term ${\cal B}[g^{n+1}](s)={s^n\over\Gamma (n+1)}$ so that
\bea
{\cal B}[{\cal O}](s)=\sum_{n=0}^\infty {c_n\over\Gamma (n)} s^{n-1}
\eea
This sum is much better behaved that the original sum and, if it converges in some region it can be used to define a function analytic in $s$, except possibly at a few singular points in the complex $s$ plane.
We can then perform an inverse map of the Borel transform (which is the Laplace transform) to complete the resummation
\bea
   {\cal S}_\theta {\cal O}=\int_0^{e^{i\theta}\infty}\,\, {\cal B}[{\cal O}](s) e^{-{s\over g}}\,\, ds
\eea
This Laplace transform is not well defined if singularities of the Borel sum lie on the contour of the $s$ integration.
Indeed, the result of the transform becomes ambiguous since it will depend on whether we go above or below the singularity.
By slowly increasing $\theta$ so that the contour of integration moves past a singularity, we find a jump in the value of the Borel resummed observable.
This is nothing but the familiar Stokes phenomenon, signaling a change in the behaviour of the asymptotics of the observable ${\cal O}$.
A pole in ${\cal B}[{\cal O}](s)$ would produce a jump proportional to $e^{-{A\over g}}$ where $A$ sets the location of the pole.
The form of the jump is reminiscent of an instanton contribution and indeed, it can be reproduced in this way.
One finds that $A$ is the classical action of an instanton.
This is a rather remarkable claim: by Borel resumming the perturbative series we can learn about non-perturbative phenomena in the problem\footnote{This is a result with a lot of history\cite{ZinnJustin:2004ib,ZinnJustin:2004cg,A1,A2,A3,Dunne:2014bca} that has recently been understood in a remarkable way \cite{Basar:2017hpr}.}. Further, it explains how to make sense of the full non-perturbative structure of the problem: the usual perturbative series should be replaced by a trans-series, which takes the form
\bea
{\cal O}=\sum_{n=0}^\infty \sigma^n {\cal O}^{(n)}(g)
\eea
where ${\cal O}^{(n)}(g)$ is the contribution of the $n$-instanton sector.
It takes the form
\bea
{\cal O}^{(n)}=e^{-{nA\over g}}\sum_{m=0}^\infty c^{(n)}_m g^m
\eea
These sums are themselves asymptotic and need to be resummed.
However the trans-series restores uniqueness to the Laplace transform: although each of the individual sums ${\cal O}^{(n)}$ jump as we pass a singularity, the complete sum ${\cal O}$ does not.
The parameter $\sigma$ is called a trans-series parameter and its role is to track instanton number.

Our goal is to determine the trans-series expansion for the giant graviton correlators we wrote down in the previous section. This will explain the structure of the large $N$ expansion for these correlators and it will make it clear what the non-perturbative contributions to the correlator are. 

\subsection{WKB Solutions}

We study the Schr\"odinger problem
\bea
\left(-{d^2\over dx^2}+N^2 Q\right)\psi =0\label{Seqn}
\eea
where
\bea
Q=\sum_{j=0} N^{-j}Q_j (x)\qquad Q_0(x)={F(x)\over G(x)}
\eea
We assume that $G(x)Q_j(x)$ are polynomials in $x$.
The small parameter $N^{-1}$ plays the role of $\hbar$. 
As usual, a turning point of the classical motion is given by a zero of $Q_0(x)$.
A simple turning point is a simple zero of $Q_0(x)$.
The poles of $Q_0(x)$ are singular points of the differential equation (\ref{Seqn}).
In the exact WKB analysis the poles and zeros of $Q_0(x)$ will play an important role.
The usual WKB ansatz
\bea
   \psi (x)=e^{\int^x dx' S(x',N)}
\eea
leads to a solution of the Schr\"odinger equation as long as $S$ solves the Riccati equation
\bea
{d S\over dx}+S^2 = N^2 Q\label{Reqn}
\eea
To solve (\ref{Reqn}), plug the ansatz
\bea
S(x,N)=\sum_{j=-1}^\infty N^{-j}S_j (x)
\eea
into (\ref{Reqn}) and equate each power of $N$ to zero.
This process yields
\bea
S_{-1}^2=Q_0\label{WKB1}
\eea
as well as the following recursion relation
\bea
S_{j+1}=-{1\over 2S_{-1}}\left( {dS_j\over dx}+\sum_{k=0}^j S_{j-k}S_k-Q_{j+2}\right)\qquad j=-1,0,1,2,\dots
\label{WKB2}
\eea
There are two possible solutions for $S_{-1}$
\bea
S_{-1}^{(\pm)}=\pm\sqrt{Q_0 (x)}
\eea
and hence there are two possible formal series solutions to the Riccati equation
\bea
S^{(\pm)}(x,N)=\sum_{j=-1}^\infty N^{-j}S_j^{(\pm)}(x)\label{seriessol}
\eea
Here ``formal series'' means formal Laurent series in $N^{-1}$.
These functions are multivalued and holomorphic on the complex plane once the zeros and poles of $Q_0(x)$ are removed. 
It is known that the series  (\ref{seriessol}) is divergent in general.
In the framework of the exact WKB analysis, the Borel resummation of the WKB solution is used to arrive at exact results.
It is useful to introduce
\bea
S_{\rm odd}(x,N)={1\over 2}\left( S^{(+)}-S^{(-)}\right)=\sum_{j=-1}^\infty N^{-j}S_{{\rm odd},j} (x)\cr
S_{\rm even}(x,N)={1\over 2}\left( S^{(+)}+S^{(-)}\right)=\sum_{j=0}^\infty N^{-j}S_{{\rm even},j} (x)
\eea
An identity that we will use below is
\bea
-{1\over 2}{d\over dx}\log S_{\rm odd}= S_{\rm even}\label{usefulident}
\eea
All that is needed to prove this identity is a simple application of (\ref{Reqn}). 
Our two possible solutions are $S=S^{(\pm)}=S_{\rm even}\pm S_{\rm odd}$.
If we use (\ref{usefulident}) we can simplify our solution as follows
\bea
   \psi_\pm (x)&=&e^{\int^x_{x_0} dx' \left(S_{\rm even}(x',N)\pm S_{\rm odd}(x',N)\right)}\cr
&=&e^{\int^x_{x_0} dx' \left(-{1\over 2}{d\over dx'}\log S_{\rm odd}(x',N)\pm S_{\rm odd}(x',N)\right)}\cr
&=& \sqrt{S_{\rm odd}(x_0,N)\over S_{\rm odd}(x,N)}e^{\pm\int^x_{x_0} dx' S_{\rm odd}(x',N)}
\eea
Since we can always multiply or divide $\psi$ by a constant\footnote{This is because (\ref{Seqn}) does not fix the normalization of $\psi$.} we can equally well take
\bea
   \psi_\pm (x)={1\over\sqrt{S_{\rm odd}(x,N)}}e^{\pm\int^x_{x_0} dx' S_{\rm odd}(x',N)}
\eea
Recall that since
\bea
S_{\rm odd}=N\sqrt{Q_0(x)}+O(1)
\eea
the WKB wave function $\psi(x)$ blows up at the simple zeroes of $Q_0(x)$.
This is nothing but the familiar breakdown of the WKB approximation at the turning points of the classical motion.

The Borel sum of the WKB solutions $\psi(x)$ have been studied in \cite{Koike}.
Consider the complex plane of parameters ${\cal C}_p$ of the Schr\"odinger equation.
The asymptotic behavior of these solutions (in $x$-space) differs in different regions of the complex ${\cal C}_p$ plane, hence the name ``parametric'' Stokes phenomenon. 
These regions are bounded in $x$-space by Stokes lines.
The Stokes line is the integral curve of ${\rm Im}(\sqrt{Q_0})dx=0$, emanating from a turning point.
Each Stokes line can either end on a singular point or on a turning point.
The complex $x$-plane is dissected into Stokes regions, each of which is bounded by Stokes curves.
The graph formed by taking the singular points and turning points as vertices and the Stokes lines as edges is called a Stokes graph.
If every edge of the Stokes graph starts on a turning point and ends on a singular point, we say that the graph is non-degenerate.
The singularities of the Borel sum (which is our main interest) will lie on these Stokes curves, so it is useful to consider them in detail.
Three Stoke's lines meet at each simple turning point.
To see this, note that close to a simple turning point at $x=a_0$ we have
\bea
\sqrt{Q_0}=\sqrt{x-a_0}R_0(x)
\eea
with $R_0(x)$ a polynomial. 
At the turning point imagine that $R_0(x=a_0)=A_0 e^{i\phi_0}$.
Change variables as follows: $(x-a_0)=re^{i\phi}$, hold $\phi$ fixed and let $r$ vary.
In this case
\bea
dx = dr e^{i\phi}
\eea
and thus
\bea
\sqrt{Q_0} dx=A_0^{3\over 2}e^{i\phi_0+i{3\over 2}\phi}dr
\eea
The condition for the Stokes line ${\rm Im}(\sqrt{Q_0})dx=0$ becomes
\bea
\phi &=& {2\over 3}(\pi n-\phi_0)
\eea
There are 3 distinct directions (for $n=0,1,2$) and hence there are 3 Stokes lines meeting at each turning point.
We will assume that all of the singular points are poles of order 2 (this is indeed the case of interest to us as we will see in the next section).
In this case, an analysis which is very similar to what we just did above, leads to the conclusion that two Stokes lines end on each singular point.
To proceed further we should specify the potential which would determine the number of singular points and turning points and hence the details of the Stokes graph.
Following this logic, we will characterize the Stokes graph for our problem in Section \ref{GGWKB}. 
To complete our review of the exact WKB method, in the next subsection we will describe the jumps in the WKB solution as we pass through a Stokes line. 

\subsection{Borel Resummation and Voros Coefficients}\label{bsandv}

Under some conditions (that we will spell out below) a suitably normalized WKB solution is Borel summable in each Stoke's region.
There are singularities on the edge of each Stoke's region that we would like to identify.
This can be accomplished by studying the Stokes phenomena of the WKB resummed solutions across the Stokes curves, since the singularities are the origin of the Stokes phenomenon.
If we take a solution to (\ref{Seqn}) and continue it along non-trivial paths in the space of parameters, we find that the solutions transform under a non-trivial monodromy group, which is another way to describe the Stokes phenomenon.
One can introduce Voros coefficients, which relate WKB solutions normalized at a turning point to WKB solutions normalized at a singular point.
The importance of the Voros coefficients follows because they capture this non-trivial monodromy group, and consequently they provide a complete characterization of the singularities of the Borel sum.

Denote the Borel sum of $\psi$ in region $D$ by $\Psi^D$.
Consider $\psi_{\pm}$ with $x_0$ chosen to be a simple turning point $x_0=a_0$.
Consider the solutions $\Psi_\pm^I$ and $\Psi_\pm^{II}$ which are the Borel sums of $\psi_\pm$ in two distinct regions $I$ and $II$.
If ${\rm Re}\left( \int_{a_0}^x \sqrt{Q_0}dx\right)  > 0$ on the boundary Stokes curve between regions $I$ and $II$, then we have\cite{Voros}
\bea
\Psi_+^I &=&\Psi_+^{II}+i\Psi_-^{II}\cr
\Psi_-^I&=&\Psi^{II}_-\label{connection}
\eea
We say that $\psi_+$ is dominant and $\psi_-$ is recessive on the Stokes curve. 
The above formulas are called connection formulas and they clearly exhibit the Stokes phenomenon for the dominant solution.
The recessive WKB solution does not have Stokes phenomena across the Stokes curves.

Consider WKB solutions normalized at a regular singular point, located at $x=r$.
In this case, $Q_0$ has a double pole at $x=r$ and, to simplify the analysis that follows, we assume that $(x-r)^2 Q_j$ for $j>0$ are holomorphic at $x=r$.
With this assumption it follows that $S_{\rm odd}$ has a simple pole at $x=r$.
To define the WKB solution at the regular singular point, we subtract this pole from $S_{\rm odd}$ and handle it analytically on its own.
To do this it is useful to introduce the expansion
\bea
\rho=\rho_0+N^{-1}\rho_1+N^{-2}\rho_2+\dots
\eea
of the function
\bea
\rho= {\rm Res}_{x=r}\,\, \sqrt{Q}\qquad\qquad{\rm where}\qquad\qquad Q=\sum_{j=0}N^{-j}Q_j
\eea
Proposition 3.6 of \cite{KawaiTakei} computes the residue (this formula assumes that $\rho$ is an even function of $N$ as explained in Appendix \ref{residue})
\bea
{\rm Res}_{x=r}\,\, S_{\rm odd}=\sigma N
\eea
where
\bea
   \sigma=\rho\sqrt{1+{1\over 4\rho^2 N^2}}
\eea
The WKB solutions normalized at the regular singular point $x=r$ are given by
\bea
\psi_\pm^{(r)}={(x-r)^{\pm\sigma N}\over\sqrt{S_{\rm odd}}}
e^{\pm\int_r^x\left(S_{\rm odd}-{\sigma N\over x-r}\right)dx}\label{singWKB}
\eea
The integrand in the above exponential is free of singularities throughout the integration domain because we have subtracted the pole at $x=r$ from $S_{\rm odd}$.
The factor $(x-r)^{\pm\sigma N}$ upfront comes from an analytic treatment of the pole contribution.
This manipulation is performed so that the integrand $S_{\rm odd}-{\sigma N\over x-r}$ is regular at the singular point, ensuring that the formula (\ref{singWKB}) is well defined.

We will now consider the recessive WKB solution at the regular singular point $x=r$.
Assume that ${\rm Re}(\rho_0)>0$. 
Then $\psi_+^{(r)}$ is recessive on any Stokes curve flowing into $x=r$.
By the connection formula, (\ref{connection}), the recessive WKB solution does not have Stoke's phenomena on the Stokes
curves. We now quote a Theorem from \cite{att}

{\vskip 0.5cm}

\noindent
{\bf Theorem 1:} Set $\tilde\psi^{(r)}_+=(x-r)^{-{1\over 2}-\sigma N}\psi_+^{(r)}$. There is a neighborhood $U$ of
$x=r$ such that $\tilde\psi^{(r)}_+$ is Borel summable in $U-\{r\}$ and $x=r$ is a removable singularity of the Borel sum
$\tilde\Psi^{(r)}_+$. Hence it is holomorphic in $U\times \{N;{\rm Re}(N)\gg 0\}$. Moreover
\bea
\tilde\Psi^{(r)}_+(r,N)=\tilde\psi^{(r)}_+(r,N)=(\sigma N)^{-{1\over 2}}
\eea
holds.

{\vskip 0.5cm}

The significance of this theorem is easy to appreciate: the factor $(x-r)^{{1\over 2}+\sigma N}$ does not admit an expansion in $1/N$. The above theorem implies that this factor appears in the WKB solution $\psi_+^{(r)}$, but this is the only non-perturbative contribution and it appears as a multiplicative factor. Indeed, as soon as it is removed (to obtain $\tilde\psi^{(r)}_+$) the result is Borel summable.
If ${\rm Re}(\rho_0)<0$ we have to exchange $+$ and $-$.

We are now ready to introduce the Voros coefficient\cite{Voros} which will play an important role in the next section.
The Voros coefficient $V_j$ describes the discrepancy between WKB solutions normalized at a turning point $a$ (denoted $\psi_\pm$) and those normalized at a singular point $b_j$ (denoted $\psi^{(b_j)}_\pm$) where $j$ specifies which singular point we consider. The definition is
\bea
  \psi_\pm^{(b_j)}=e^{\pm V_j}\psi_\pm
\eea
This completes our review of the exact WKB solutions. In the next section we apply the method to the hypergeometric differential equation.

\section{Application of Exact WKB to Giant Graviton Correlators}\label{GGWKB}

We have seen in Section \ref{correlators} that the normalized three point function of giant graviton correlators can be expressed in terms of the hypergeometric function ${}_2F_1(a,b;c;1)$.
In this section we will map the hypergeometric differential equation into a Schr\"odinger equation.
We can then apply the results of the previous section to perform an exact WKB analysis.
Under the mapping to the Schr\"odinger equation, $1/N$ maps to $\hbar$ so that the semi-classical expansion for the Schr\"odinger equation is the $1/N$ expansion of our correlators. This implies that through this map we are able to understand the structure of the ${1\over N}$ expansion in this large $N$ but non-planar limit.

\subsection{Mapping to the Schr\"odinger Equation}

The hypergeometric differential equation is
\bea
   x(1-x){d^2 w\over dx^2}+(c-(a+b+1)x){dw\over dx}-abw=0\label{hgde}
\eea
Notice that it has regular singular points at $b_0=0$, $b_1=1$ and $b_2=\infty$.
The parameters of the hypergeometric function are
\bea
     a={1\over 2}+\alpha N\qquad b={1\over 2}+\beta N\qquad c=1+\gamma N
\eea
where $N$ is taken to be large.
This particular parametrization of $a,b,c$ follows \cite{att} and will simplify many of the formulas that follow.
Introduce the wave function $\psi$ as follows
\bea
   \psi = x^{{1\over 2}(1+\gamma N)}(1-x)^{{1\over 2}(1+(\alpha+\beta-\gamma)N)}w
\eea
Plugging this into (\ref{hgde}) we find that $\psi$ obeys the following Schr\"odinger equation
\bea
   \left( -{1\over N^2}{d^2\over dx^2}+ Q(x)\right)\psi =0
\eea
where
\bea
   Q(x)=Q_0(x)+N^{-2}Q_2 (x)
\eea
\bea
   Q_0(x)={(\alpha-\beta)^2 x^2 +2(2\alpha\beta-\alpha\gamma-\beta\gamma)x+\gamma^2\over 4x^2 (x-1)^2}
   \label{Q0}
\eea
\bea
Q_2(x)=-{x^2-x+1\over 4x^2 (x-1)^2}
\eea
An important and non-trivial feature of this mapping is that we see that ${1\over N}$ plays the role of $\hbar$.
This is in complete agreement with the usual holographic dictionary between CFT parameters and the parameters of the dual gravity, so one may wonder if this Schr\"odinger equation has a natural gravitational origin. 
We will not explore this possibility in this article.
Since $Q_0(x)$ is quadratic, there are two turning points $\{ a_0,a_1\}$, given by the zeros of the numerator of (\ref{Q0}).

To properly define the coefficient $S_{-1}=\sqrt{Q_0}$ of the WKB solution, we need to explain what branch of $\sqrt{Q_0}$ we use. The branch cut runs between the two turning points avoiding the singular points $b_k$. The branch we use is specified by choosing 
\bea
  \sqrt{Q_0}\sim{\gamma\over 2x}\qquad &{\rm at}&\qquad x=0\cr
  \sqrt{Q_0}\sim{\alpha+\beta-\gamma\over 2(x-1)}\qquad &{\rm at}&\qquad x=1\cr
  \sqrt{Q_0}\sim{\beta-\alpha\over 2x}\qquad &{\rm at}&\qquad x=\infty
\eea

\subsection{Stokes Graph}

The (complexified) position space with coordinate $x$ on which the wave function is defined is divided up into regions by the Stokes graph.
The Stokes graph of (\ref{hgde}) is the graph drawn on the sphere with vertices given by the turning points $\{ a_0,a_1\}$ and the regular singular points $\{b_0,b_1,b_2\}$ and edges given by Stokes lines.
The WKB solutions jump discontinuously across the Stokes lines, which is the usual Stokes phenomenon.
The Stokes graph of (\ref{hgde}) is well understood\cite{Tanda}.
Since this will be needed in what follows, we review the relevant results of \cite{Tanda} in this section.

The topology of the Stokes graph can change depending on the values of the parameters appearing in the potential.
We imagine that the parameters $\alpha$, $\beta$ and $\gamma$ are arbitrary complex numbers taking values on the Riemann sphere ${\cal C}_p$.
We can divide this space up into regions, such that the topology of the Stokes graph is fixed in each region.
Towards this end, introduce the following three sets
\bea
E_0&=&\{ (\alpha,\beta,\gamma)\in {\mathbb C}^3 |\alpha\beta\gamma (\alpha-\beta)(\alpha-\gamma)(\beta-\gamma)
(\alpha+\beta-\gamma)=0\}\cr\cr
E_1&=&\{ (\alpha,\beta,\gamma)\in {\mathbb C}^3 |{\rm Re}(\alpha){\rm Re}(\beta){\rm Re}(\gamma -\alpha)
{\rm Re}(\gamma -\beta)=0\}\cr\cr
E_2&=&\{ (\alpha,\beta,\gamma)\in {\mathbb C}^3 |{\rm Re}(\alpha-\beta){\rm Re}(\alpha+\beta-\gamma){\rm Re}(\gamma)=0\}
\eea
To get some insight into the definition of the above open sets, note that
\bea
\alpha\beta\gamma (\alpha-\beta)(\alpha-\gamma)(\beta-\gamma)(\alpha+\beta-\gamma)\ne 0
\eea
is the condition that there are two distinct turning points and further that neither turning points coincides with a singular point.
The conditions
\bea
{\rm Re}(\alpha){\rm Re}(\beta){\rm Re}(\gamma -\alpha)
{\rm Re}(\gamma -\beta)\ne 0\ne {\rm Re}(\alpha-\beta){\rm Re}(\alpha+\beta-\gamma){\rm Re}(\gamma)
\eea
ensure that there is no Stokes curve connecting distinct turning points (the first condition) or the same turning point (the second condition). 
If turning points are connected by a Stokes curve, the Stokes geometry is said to be degenerate. 
The conditions under which the Stokes graph is degenerate is summarized in the following theorem

{\vskip 0.5cm}

\noindent
{\bf Theorem 2:} Assume that $(\alpha,\beta,\gamma)$ is not contained in $E_0$.
(i) If two distinct turning points $a_0$ and $a_1$ are connected by a Stokes curve, then $(\alpha,\beta,\gamma)$ belong to $E_1$. Conversely, if $(\alpha,\beta,\gamma)$ is contained in $E_1-E_2$, the Stokes geometry of (\ref{hgde}) has a Stokes curve which connects two distinct turning points $a_0$ and $a_1$. (ii) If a Stokes curve forms a closed curve with a single turning point as the base point, then $(\alpha,\beta,\gamma)$ belongs to $E_2$. Conversely if $(\alpha,\beta,\gamma)$ is contained in $E_2-E_1$, the Stokes geometry of (\ref{hgde}) has a Stokes curve which forms a closed path with a turning point as the base point.

{\vskip 0.5cm}

To proceed further we need to define the following sets of parameters
\bea
\omega_1&=&\{(\alpha,\beta,\gamma)\in{\mathbb C}^3|0<{\rm Re}(\alpha)<{\rm Re}(\gamma)<{\rm Re}(\beta)\}\cr\cr
\omega_2&=&\{(\alpha,\beta,\gamma)\in{\mathbb C}^3|0<{\rm Re}(\alpha)<{\rm Re}(\beta)<{\rm Re}(\gamma)<
{\rm Re}(\alpha)+{\rm Re}(\beta)\}\cr\cr
\omega_3&=&\{(\alpha,\beta,\gamma)\in{\mathbb C}^3|0<{\rm Re}(\gamma)<{\rm Re}(\alpha)<{\rm Re}(\beta)\}\cr\cr
\omega_4&=&\{(\alpha,\beta,\gamma)\in{\mathbb C}^3|0<{\rm Re}(\gamma)<{\rm Re}(\alpha)+{\rm Re}(\beta)<
{\rm Re}(\beta)\}
\eea
 as well as the involutions
\bea
\iota_0:(\alpha,\beta,\gamma)&\to& (-\alpha,-\beta,-\gamma)\cr\cr
\iota_1:(\alpha,\beta,\gamma)&\to& (\gamma-\beta,\gamma-\alpha,\gamma)\cr\cr
\iota_2:(\alpha,\beta,\gamma)&\to& (\beta,\alpha,\gamma)
\eea
The relevance of these involutions follows because they are symmetries of the potential $Q(x)$, so parameters related by the involution give the same solution. Let $G$ be the group generated by $\iota_j$ $j=0,1,2$. $G$ is then a discrete group of symmetries of $Q$.
Define the open subsets
\bea
\Pi_h=\bigcup\limits_{r\in G}r(\omega_h)\qquad h=1,2,3,4
\eea
The union of the $\Pi_h$ covers most of ${\mathbb C}^3$:
\bea
\bigcup\limits_{h=1}^4 \Pi_h ={\mathbb C}^3 - {\cal U}
\eea 
where
\bea
{\cal U}&=&\{(\alpha,\beta,\gamma)|{\rm Re}(\alpha){\rm Re}(\beta){\rm Re}(\gamma)\cr
&\times& {\rm Re}(\gamma-\alpha)
{\rm Re}(\gamma-\beta){\rm Re}(\alpha-\beta){\rm Re}(\alpha+\beta-\gamma)=0\}
\eea
The topological structure of the Stokes graph can be summarized by the triple of integers $(n_0,n_1,n_2)$ where $n_j$ counts how many Stokes curves flow into the regular singular point $b_j$.
The topological structure of the Stokes graph is summarized in the following theorem

{\vskip 0.5cm}

\noindent
{\bf Theorem 3:} Let $\hat n=(n_0,n_1,n_2)$ denote the order sequences of the Stokes graph with parameters $(\alpha,\beta,\gamma)$. If $(\alpha,\beta,\gamma)\in\Pi_1$ then $\hat{n}=(2,2,2)$. If $(\alpha,\beta,\gamma)\in\Pi_2$ then $\hat{n}=(4,1,1)$. If $(\alpha,\beta,\gamma)\in\Pi_3$ then $\hat{n}=(1,4,1)$. If $(\alpha,\beta,\gamma)\in\Pi_4$ then $\hat{n}=(1,1,4)$.

{\vskip 0.5cm}

\subsection{Voros coefficients}

From the discussion in Section \ref{bsandv}, it is straightforwards to see that the Voros coefficient accounting for the discrepancy between the WKB solutions normalized at turning point $a$ and those normalized at singular point $b_k$ are given by
\bea
   V_k(\alpha,\beta,\gamma)=\int_{b_k}^a (S_{\rm odd}-NS_{-1})dx
\eea
The residues of $S_{\rm odd}$ and $NS_{-1}$ at the singular points coincide which implies that the $V_k(\alpha,\beta,\gamma)$ are well defined and that we can develop a formal power series in $N^{-1}$. The explicit power series are\cite{AokiandTanda}
\bea
V_0&=&-{1\over 2}\sum_{n=2}^\infty {B_n N^{1-n}\over n(n-1)}\left[(1-2^{1-n})\left({1\over\alpha^{n-1}}
+{1\over\beta^{n-1}}+{1\over (\gamma-\alpha)^{n-1}}+{1\over (\gamma-\beta)^{n-1}}\right)+{2\over\gamma^{n-1}}\right]\cr
V_1&=&{1\over 2}\sum_{n=2}^\infty {B_n N^{1-n}\over n(n-1)}\left[(1-2^{1-n})\left({1\over\alpha^{n-1}}
+{1\over\beta^{n-1}}-{1\over (\gamma-\alpha)^{n-1}}-{1\over (\gamma-\beta)^{n-1}}\right)
+{2\over(\alpha+\beta-\gamma)^{n-1}}\right]\cr
V_2&=&{1\over 2}\sum_{n=2}^\infty {B_n N^{1-n}\over n(n-1)}\left[(1-2^{1-n})\left({1\over\alpha^{n-1}}
-{1\over\beta^{n-1}}-{1\over (\gamma-\alpha)^{n-1}}+{1\over (\gamma-\beta)^{n-1}}\right)
-{2\over(\beta-\alpha)^{n-1}}\right]\nonumber
\eea
where $B_n$ are the Bernoulli numbers defined by
\bea
{te^t\over e^t-1}=\sum_{n=0}^\infty {B_n\over n!}t^n
\eea
Noting the asymptotic growth of the Bernoulli numbers
\bea
B_{2k}\sim 4\left({k\over\pi e}\right)^{2k}\sqrt{\pi k}
\eea
it is clear that the series expansions given above are asymptotic series.
The Borel transforms of the above series are well defined and are given by\cite{Tanda}
\bea
{\cal B}[V_0](y)&=&-{1\over 4}\left[g_1(\alpha;y)+g_1(\beta;y)+g_1(\gamma-\alpha;y)+g_1(\gamma-\beta;y)\right]
+g_0(\gamma;y)\cr
{\cal B}[V_1](y)&=&{1\over 4}\left[-g_1(\alpha;y)-g_1(\beta;y)+g_1(\gamma-\alpha;y)+g_1(\gamma-\beta;y)\right]
+g_0(\alpha+\beta-\gamma;y)\cr
{\cal B}[V_2](y)&=&{1\over 4}\left[-g_1(\alpha;y)+g_1(\beta;y)+g_1(\gamma-\alpha;y)-g_1(\gamma-\beta;y)\right]
-g_0(\beta-\alpha;y)
\eea
where
\bea
   g_0(t;y)&=&{1\over y}\left({1\over e^{y\over t}-1}+{1\over 2}-{t\over y}\right)\cr
   g_1(t;y)&=&{1\over e^{y\over 2t}-1}+{1\over e^{y\over 2t}+1}-{2t\over y}
\eea
These functions have singularities which signals both the Stokes phenomenon of the asymptotic series and non-perturbative behaviour in the field theory. Both functions have simple poles at $y=2tm\pi i$ with $m$ any nonzero integer. The residues of these poles are
\bea
  \res_{y=2tm\pi i}g_0(t;y)={1\over 2\pi mi}\qquad\qquad\qquad
  \res_{y=2tm\pi i}g_1(t;y)={(-1)^m\over \pi mi}
\eea
The results can be used to compute alien derivatives and the Stokes automorphims for the WKB solutions. The interested reader can find a clear readable account in \cite{Tanda}.

\subsection{Trans-series Expansion of Giant Graviton Three Point Function}

Our primary goal in this section is to relate the Borel sum of the WKB solution to ${}_2F_1(a,b;c;x)$ near $x=1$.
Our approach is based on the study \cite{att} which established the relationship between the Borel sum of the WKB solution
to ${}_2F_1(a,b;c;x)$ near $x=0$. 
Specifically we will study the leading contribution to the WKB solution and show that it reproduces the leading behavior of the correlator, which is non-perturbative in $1/N$. The relation between the hypergeometric function and the WKB solution normalized at $x=0$  is\cite{att}
\bea
 F(\frac{1}{2}+\alpha N,\frac{1}{2}+\beta N,1+\gamma N;x)=\sqrt{\frac{\gamma}{2}}N^{1/2}e^{-N h_0}x^{-\frac{1}{2}(1+\gamma N)}(1-x)^{-\frac{1}{2}-\frac{\alpha+\beta-\gamma}{2}N}\cr
 \times \frac{1}{\sqrt{S_{odd}}}\exp\left[N\int_{0}^x S_{-1}dx+\int_0^x(S_{odd}-N S_{-1})dx\right]
\eea
Note that we have normalized the wave function using the value of the hypergeometric function at $x=0$, i.e.
$F(a,b;c;x=0)=1$. For the leading order at large N, we only need the first integral in the exponential term on the right hand side
\bea
 F(\frac{1}{2}+\alpha N,\frac{1}{2}+\beta N,1+\gamma N;x)&\approx&\sqrt{\frac{\gamma}{2}}N^{1/2}e^{-N h_0}x^{-\frac{1}{2}(1+\gamma N)}(1-x)^{-\frac{1}{2}-\frac{\alpha+\beta-\gamma}{2}N}\cr
 &\times& \frac{1}{\sqrt{S_{odd}}}\exp\left[N\int_{0}^x S_{-1}dx\right],
 \eea
 where,
\bea
  S_{-1}&=&\sqrt{Q_0(x)},\\
   Q_0(x)&=&{(\alpha-\beta)^2 x^2 +2(2\alpha\beta-\alpha\gamma-\beta\gamma)x+\gamma^2\over 4x^2 (x-1)^2}.
   \label{Q0}\\\nn
   h_0&=&\frac{1}{4}(\alpha \ln\alpha^2+\beta\ln\beta^2+(\gamma-\alpha)\ln(\alpha-\gamma)^2+(\gamma-\beta)\ln(\beta-\gamma)^2+2\gamma\ln\gamma^2)\\
\eea
It is straight forwards to see that
\bea\label{exp}
&&4\int S_{-1} dx=T_1+T_2+T_3
\eea
where
\bea
&&T_1=p_3 \log \left(\frac{2p_2-2p_3\sqrt{p_1(x)}+2xp_3^2}{2 p_2+2p_3\sqrt{p_1(x)}+2xp_3^2}\right)\quad
T_2=\gamma  \log \left(\frac{2 \gamma ^2-2 \gamma  \sqrt{p_1(x)}+2 x p_2}{2 \gamma ^2+2 \gamma  \sqrt{p_1(x)}+2 x p_2}\right)\cr\cr
&&T_3=(-\alpha -\beta +\gamma ) \log \left(\frac{2 p_2 +2 \gamma ^2+2 (-\alpha -\beta +\gamma ) \sqrt{p_1(x)}+x \left(2 p_2+2 p_3^2\right)}{2 p_2+2 \gamma ^2-2 (-\alpha -\beta +\gamma ) \sqrt{p_1(x)}+x \left(2 p_2+2 p_3^2\right)}\right)
\eea
and
\bea
p_1(x)&=&\gamma ^2+x^2 (\alpha -\beta )^2+2 x (2 \alpha  \beta -\alpha  \gamma -\beta  \gamma )\cr
p_2&=&2 \alpha  \beta -\alpha  \gamma -\beta  \gamma \qquad
p_3=\beta - \alpha
\eea

Now consider the expansion of $F(\frac{1}{2}+\alpha N,\frac{1}{2}+\beta N,1+\gamma N;x)$ about $x=1$.
$S_{odd}$ can be expanded about $x=1$ as
\bea
S_{odd}=\frac{c_{-1}}{1-x}+c_0+c_1(1-x)+...,
\eea
As $x\rightarrow 1$ we have
\bea
\frac{1}{\sqrt{S_{odd}}}\sim \frac{\sqrt{1-x}}{\sqrt{c_{-1}+c_0(1-x)+c_{1}(1-x)^2+...}}\sim \frac{\sqrt{1-x}}{\sqrt{c_{-1}}}=(\frac{1}{2}\g N)^{-1/2}(1-x)^{1/2}
\eea
\bea\label{F}
F(\frac{1}{2}+\alpha N,\frac{1}{2}+\beta N,1+\gamma N;x)\rightarrow e^{-N h_0}(1-x)^{-\frac{\a+\b-\g}{2}N}\exp\left[N\int_{a_0}^{x}\sqrt{Q_0}dx\right]
\eea
and
\bea
T_1&\rightarrow& p_3\ln\frac{\b(\g-\b)}{\a(\g-\a)}=(\b-\a)\ln\frac{\b(\g-\b)}{\a(\g-\a)}\\
T_2&\rightarrow& \g\ln\frac{\a\b}{(\g-\a)(\g-\b)}
\eea
\bea
T_3&\rightarrow& (\gamma-\alpha-\beta)\ln\frac{4(\gamma-\alpha-\beta)^2}{A}-2(\gamma-\alpha-\beta)\ln\epsilon+O(\epsilon)\\
    &=&(\gamma-\alpha-\beta)\ln\frac{(\gamma-\alpha-\beta)^4}{\alpha\beta(\gamma-\alpha)(\gamma-\beta)}-\ln\epsilon^{2(\gamma-\alpha-\beta)}+O(\epsilon)
\eea
With these results we find that as $x\rightarrow 1$ we have
\bea
\int_{a_0}^{x}\sqrt{Q_0}dx=\frac{1}{4}(T_1+T_2+T_3)=\ln(1-x)^{-1/2(\g-\a-\b)}+h_1,
\eea
where
\bea
h_1=(\g-\a-\b)\ln(\g-\a-\b)
+\frac{1}{4}\a\ln\a^2+\frac{1}{4}\b\ln\b^2\\+\frac{1}{4}(\a-\g)\ln(\g-\a)^2+\frac{1}{4}(\b-\g)\ln(\g-\b)^2.
\eea
This finally gives
\bea
F(\frac{1}{2}+\alpha N,\frac{1}{2}+\beta N,1+\gamma N;x=1)=e^{N(h_1- h_0)}
\eea
where
\bea
\nn
h_1- h_0=\frac{1}{2}(\a-\g)\ln(\g-\a)^2+\frac{1}{2}(\b-\g)\ln(\g-\b)^2+(\g-\a-\b)\ln(\g-\a-\b)+\g\ln\g\\
\nn
=(\a-\g)\ln(\g-\a)+(\b-\g)\ln(\g-\b)+(\g-\a-\b)\ln(\g-\a-\b)+\g\ln\g.
\eea

We can now use these results to obtain the leading behavior of the normalized correlation function of three giant gravitons.   
For the $\langle O_{A_{J_1}}O_{A_{J_2}}O_{A_{J}}\rangle$ correlation function we have, at large $N$
$\a= -j_1, ~\b= j_2, \g=1+\a=1-j_1 $. Using these parameter values we find
\bea\nn
h_1- h_0&=&(\b-\a-1)\ln(1+\a-\b)+(1-\b)\ln(1-\b)+(1+\a)\ln(1+\a)\\\nn
&=&(j_1+j_2-1)\ln(1-j_1-j_2)+(1-j_2)\ln(1-j_2)+(1-j_1)\ln(1-j_1)\\\nn
&\sim& (j_1+j_2-1)(-j_1-j_2)+(1-j_2)(-j_2)+(1-j_1)(-j_1)\\
&=&-2j_1j_2.
\eea
where we have expanded assuming small $j_1$ and $j_2$ in the second last line.
Terms cubic in $j_1,j_2$ were neglected.
Thus, we finally obtain
 \bea
 F(\frac{1}{2}+\alpha N,\frac{1}{2}+\beta N,1+\gamma N;x=1)= e^{-2Nj_1j_2}
 \eea
for the leading contribution at large $N$.
This is in complete agreement with (\ref{NPcorr}).
We can also consider the $\langle O_{S_{J_1}}O_{S_{J_2}}O_{S_{J}}\rangle$ correlation function of three dual giant gravitons.
In this case, again at large $N$, we have $\a= j_1, ~\b= -j_2, \g=1+\a=1+j_1 $. 
The only difference to the case we have just considered is that $j_1$ and $j_2$ change sign. 
Thus, the answer for $h_1-h_0$ is unchanged. In this case however, we expect the correlator to behave as $e^{2Nj_1j_2}$.
To see how the transition from the exponentially dying to exponentially growing solution occurs, recall that $S_{-1}=\sqrt{Q_0}$ is only defined up to a sign. Taking the $+$ sign reproduces the correlation function of three giant gravitons. Taking the $-$ sign reproduces the correlation function of three dual giant gravitons.
This is a convincing demonstration that the WKB wave function is indeed reproducing the giant graviton correlators and, since we know the form of the semi-classical expansion of the WKB wave function, it immediately gives the form of the $1/N$ expansion of the giant graviton correlation function. Using an obvious notation, we have
\bea
\langle O_{J_1}O_{J_2}O^\dagger_{J_1+J_2}\rangle =
e^{\pm j_1 j_2 N}\sum_{n=1}^\infty c_n N^{-n}
\eea
In terms of $g_s={1\over N}$ this can be written as
\bea
\langle O_{J_1}O_{J_2}O^\dagger_{J_1+J_2}\rangle =
e^{\pm{j_1 j_2\over g_s}}\sum_{n=1}^\infty c_n g_s^{n}
\eea
This clearly exhibits the non-analyticity of the expansion and shows that the transeries expansion of the giant graviton correlators is particularly simple.

Lets now consider the Stokes graph relevant for the giant graviton problem. Start from the parameters
\bea
(\a,\b,\g)=(-j_1,j_2,1-j_1)
\eea
and apply the involution $\iota_1$ to obtain
\bea
(\a,\b,\g)=(1-j_1-j_2,1,1-j_1)
\eea
Clearly we have $0<{\rm Re}(\a)<{\rm Re}(\g)<{\rm Re}(\b)$ so that the parameters relevant to the giant graviton problem are in $\Pi_1$. The boundaries of this region have a transparent physical interpretation. The boundary at which ${\rm Re}(\alpha)=0$ corresponds to $j_1+j_2=1$ so that $O^\dagger_{S_{J_1+J_2}}$ is a maximal giant graviton. The boundary ${\rm Re}(\a)={\rm Re}(\g)$ corresponds to $j_2=0$ so that $O_{S_{J_2}}$ reduces to a point graviton. Finally, the boundary ${\rm Re}(\b)={\rm Re}(\g)$ corresponds to $j_1=0$ so that now it is $O_{S_{J_1}}$ that reduces to a point graviton.

Finally, note that for the dual giant graviton we have
\bea
(\a,\b,\g)=(j_1,-j_2,1+j_1)
\eea
First applying $\iota_2$ and then $\iota_1$ we find 
\bea
(\a,\b,\g)=(1,1+j_1+j_2,1+j_1)
\eea
so that these parameters again lie in $\Pi_1$. There is an interesting difference between giant gravitons and dual giant gravitons: the angular momentum of a giant graviton is limited by $N$ but there is no limit on the angular momentum of the dual giant graviton. It is interesting then to again ask about the boundaries of $\Pi_1$. The boundary ${\rm Re}(\b)={\rm Re}(\g)$ corresponds to $j_2=0$ so that $O_{A_{J_2}}$ reduces to a point graviton and the boundary ${\rm Re}(\g)={\rm Re}(\a)$ corresponds to $j_1=0$ so that $O_{A_{J_1}}$ reduces to a point graviton. Finally, the boundary ${\rm Re}(\a)=0$ is never realized.

What this analysis proves is that after removing the overall non-perturbative factor, the correlator does admit a $1/N$ expansion and the resulting series is Borel summable everywhere in the domain of allowed values for the giant graviton momenta $J_1,J_2$. 
The Stokes lines demarcating the region in which the correlator is Borel summable, correspond to the limits in which giant gravitons tend to their point like limit or become maximal. These are precisely the values of $J$ at which we transition from one partial representation (as a brane) to another partial representation (as a string) of the holographic dual to the CFT operator.

\section{Discussion}\label{discuss}

In this article we have studied the structure of the large $N$ expansion in certain large $N$ but non-planar limits of ${\cal N}=4$ super Yang-Mills theory.
We have considered three point extremal correlation functions of ${1\over 2}$ BPS operators that have a dimension of order $N$. 
They are interpreted as giant gravitons and dual giant gravitons in the dual gravitational description.
These correlators do not receive 't Hooft coupling corrections and have been computed exactly to all orders in $1/N$.
It is thus somewhat natural to expect that much can be learned about non-perturbative phenomena in the large $N$ expansion from these correlators.

For three point functions of operators dual to giant gravitons, that have a dimension of order $N$, we have argued that the large $N$ expansion takes the form ($j_1$ and $j_2$ are both of order 1)
\bea
\langle O_{A_{N j_1}}O_{A_{N j_2}}O_{A_{N(j_1+j_2)}}\rangle =e^{-Nj_1 j_2}\sum_{n=1}^\infty c_n N^{-n}
\label{e3pn}
\eea
There is a similar result for the three point function of three dual giant gravitons.
This is a non-perturbative result since the exponential factor can not be expanded in $1/N$. The exact correlator is
therefore the product of a non-perturbative factor with a perturbative factor.  
The non-perturbative contribution has been explained using instanton configurations \cite{Hirano:2018xmh} in the tiny graviton theory.
The tiny graviton matrix model\cite{SheikhJabbari:2004ik} is a proposal for the discrete lightcone quantization of IIB string
theory on the maximally supersymmetric ten-dimensional plane wave. 
It is a matrix quantum mechanics of $J\times J$ matrices.
Instanton solutions of the model were studied in \cite{Hirano:2018xmh}. 
Remarkably, the instanton induced splitting and joining of giant gravitons developed in \cite{Hirano:2018xmh} is in complete agreement with the correlation functions of normalized Schur polynomials.
It would be interesting to evaluate ${1\over N}$ corrections in the tiny graviton theory and see if the subleading terms in
(\ref{e3pn}) can be reproduced.
Further, we have argued that after the non-perturbative factor in (\ref{e3pn}) is removed, the correlator admits a $1/N$ expansion that is Borel summable. 
The Stokes lines across which the Borel sum has a discontinuity occur at the values of $J$ at which we transition from one partial representation (as a brane) to another partial representation (as a string) of the holographic dual to the CFT operator.
As we described in the introduction, the dual to a single CFT operator transitions through different physical descriptions (particles, strings and branes) as the dimension of the operator is varied.
We vary the dimension smoothly and the variation of the CFT correlators is not singular at all.
The reorganization of the $1/N$ expansion however changes discontinuously as a consequence of moving between the different descriptions.
Our results suggest that the transition between different representations are accompanied by a Stokes phenomena and it is natural to expect that resurgence will play a role when these different partial representations are combined into a single coherent description.
Our analysis has explicitly assumed that the parameters $J_i$ are of order $N$. 
It would be interesting to generalize our analysis and to follow the description all the way down to $J_i\sim 1$.
In this case it may be possible to continue the WKB analysis through the Stokes line and verify if there is a transition
from a brane-like description to a string description.

The suggestion that Stokes lines might separate different physical descriptions of a single system, with different descriptions realized by varying parameters in the problem, might be considered in other contexts too.   
In the recent article \cite{Banerjee:2018cjx} non–perturbative interpolating functions to probe the physics of the cusp and twist-two anomalous dimensions were constructed.
Finite $N$ ${\cal N} = 4$ SYM is expected to be S-duality invariant. 
To probe this physics, \cite{Banerjee:2018cjx} accounts for both non–planar and instanton contributions by constructing modular invariant interpolating functions. 
At the two ends the anomalous dimensions scale as $\sim\sqrt\lambda$ or $\sim\lambda^{1\over 4}$.
The cusp anomalous dimension emerges in the large spin limit while the twist two operators are considered in the small spin limit.
The two descriptions are long spinning ``spiky'' strings versus small circular strings.
It again seems natural to guess there is a Stokes line separating these distinct saddles that must be crossed as the spin varies. 

In the usual planar limit, the $1/N$ expansion has a compelling physical interpretation \cite{tHooft:1973alw}.
The Feynman diagram expansion is in terms of ribbon graphs.
The small parameter of the expansion is ${1\over N^2}$ and the power of $N$ multiplying a given term has a nice interpretation as the genus of the worldsheet corresponding to the ribbon graph. 
Is there an interpretation for the series multiplying the non-perturbative term in (\ref{e3pn})?
An approach towards this problem is suggested by recalling that the giant gravitons are spherical D3 branes and their 
excitations can be described in terms of open strings.
The non-perturbative factor in (\ref{e3pn}) is naturally associated to the spherical D3 brane\footnote{It is naturally interpreted as the exponential of the D3 brane action. The D3 action is multiplied by the tension of a D-brane which behaves like $\sim{1\over g_s}\sim N$ since $g_s\sim{1\over N}$.}, while the series multiplying this factor is naturally associated to the open string theory living on the giant graviton.
From this point of view, powers of $N$ would be associated  to the genus of world sheets for the open strings. 
This provides a natural explanation of why the perturbative factor is a series in ${1\over N}$ and not ${1\over N^2}$.
We hope to further develop this point of view.

The mapping to the Schr\"odinger equation has allowed us to find the form of the ${1\over N}$ expansion and to argue that the asymptotic series coming from the ${1\over N}$ expansion of giant graviton correlators is Borel summable. 
If all that we are interested in is the form of the ${1\over N}$ expansion, then because our correlators are ratios of $\Gamma (\cdot)$ functions, we can simply use the known asymptotic expansion
\bea
\Gamma (z)=e^{-z}z^{z-{1\over 2}}\sqrt{2\pi}\left(1+{1\over 12z}+{1\over 288z^2}+\cdots\right)
\eea
Plugging this into the giant graviton correlators we easily recover the form we obtained from the WKB analysis.
Repeating this logic, we find the following form for the expansion of the general giant and dual giant extremal correlators
\bea
\langle O_{A_{J_1}}\cdots O_{A_{J_k}}O^\dagger_{A_{J_1+J_2+\cdots+A_k}}\rangle=
e^{-N\sum_{i>j=1}^{k-1}j_i j_i}\sum_{n=0}^\infty c_n N^{-n+k-1}
\eea
\bea
\langle O_{S_{J_1}}\cdots O_{S_{J_k}}O^\dagger_{S_{J_1+J_2+\cdots+A_k}}\rangle=
e^{N\sum_{i>j=1}^{k-1}j_i j_i}\sum_{n=0}^\infty c_n N^{-n+k-1}
\eea
Given our experience with the three point functions, we conjecture that these series will be Borel summable in the 
physically allowed range of parameters, which is $j_i>0$ for $i=1,2,\cdots,k$ and $j_1+\cdots j_k\le 1$ for the
giant gravitons.  It would be interesting to explicitly prove this.

We have considered general extremal correlators between giant gravitons and between dual giant gravitons.
The complete class of extremal correlation functions of Schur polynomials is much more general. 
It would also be interesting to study correlators involving operators with a dimension of order $N^2$.
These would have a gravitational interpretation in terms of physics in an LLM geometry\cite{Lin:2004nb}, so that one is probing a back reacted version of the AdS$_5\times$S$^5$ spacetime.
It is interesting to ask what the structure of the large $N$ expansion in this case is?
Once again the extremal correlation functions can be evaluated exactly.
In simple examples \cite{deMelloKoch:2009jc} for well chosen backgrounds, the only effect on the extremal correlators is a renormalization $N\to N_{\rm eff}$.
The perturbative expansion in this LLM background becomes an expansion in ${1\over N_{\rm eff}^2}$, which suggests that the closed string coupling constant $g_s$ has been renormalized.
This same effect has also been observed beyond the half-BPS sector \cite{Koch:2016jnm,deMelloKoch:2018tlb,deMelloKoch:2018ert,Kim:2018gwx}.
Does this renormalization of $N$ persist when non-perturbative corrections are considered?
This could be probed by studying giant graviton correlators in the LLM background.
If the closed string coupling is renormalized, then the tension of the D3-brane $\sim {1\over g_s}$ should be 
renormalized and we do expect the renormalization of $N$ to persist. 

The true power of resurgence only comes into play when we have many non-perturbative sectors as well as a perturbative
sector.
Resurgence then relates the series in these different sectors (see \cite{Basar:2017hpr} for recent results and references).
Extremal giant graviton and dual giant graviton correlators have only a single sector.
Further, the form of our extremal correlators makes it likely that we need to go beyond the half-BPS sector for
correlators that have more than a single sector.

Finally, it maybe worth rexamining the ``analytic bootstrap'' for the exact WKB method, formulated in \cite{Voros}. 
The method considers WKB periods, which are (Borel resummed) perturbative series in $\hbar$. 
These periods are determined by their classical limit and their discontinuity structure, which is encoded in the Stokes automorphisms. 
This data defines a Riemann-Hilbert problem, which can be solved in terms of a TBA-like system, uncovering a remarkable correspondence between ordinary differential equations and integrable models\cite{Dorey:1998pt,Ito:2018eon}.
It would be interesting to revisit the analytic bootstrap, considering the role of discontinuities associated with the
parametric Stokes phenomenon.

{\vskip 0.5cm}
\noindent
\begin{centerline} 
{\bf Acknowledgements}
\end{centerline} 

This work is supported by the South African Research Chairs initiative of the Department of Science and Technology 
and the National Research Foundation.
J.-H.Huang is supported by the Natural Science Foundation of Guangdong Province (No.2016A030313444).
We are grateful for useful discussions to Aritra Banerjee, Warren Carlson and Shinji Hirano.
We are also very grateful to Mika Tanda for patient correspondence on the WKB analysis of the hypergeometric function. 

\appendix

\section{Residue of $S_{\rm odd}$}\label{residue}

To determine the WKB solution normalized at the regular simgular point $x=1$, we need to evaluate the resdiue of $S_{\rm odd}$ at $x=1$. This has been carried out in detail in \cite{KawaiTakei} - see Proposition 3.6. Here we will give a quick summary of the argument, both to make the paper self contained and to stress the differences between our case and the case of \cite{KawaiTakei}. We will make use of the Ricatti equation (\ref{Reqn}) which we repeat for convenience
\bea
{d S\over dx}+S^2 = N^2 Q
\eea
First, note that
\bea
  \sqrt{Q(x)}={\sqrt{F(x)}\over 2x(x-1)}
\eea
with
\bea
F(x)=(\alpha-\beta)^2 x^2 +2(2\alpha\beta-\alpha\gamma-\beta\gamma)x+\gamma^2 - N^{-2}\left(x^2-x+1\right)
\eea
It is trivial to see that
\bea
{\rm Res}_{x=1}\,\, \sqrt{Q(x)}={\sqrt{(\alpha-\beta)^2+2(2\alpha\beta-\alpha\gamma-\beta\gamma)+\gamma^2 - N^{-2}}\over 2}\equiv \rho(N)
\eea
The formal solution to the Riccati equation is given by a power sum
\bea
  S(x,N)=\sum_{j=-1}^\infty S_j (x) N^{-j}\label{powerser}
\eea
Plugging this sum into the Riccati equation leads to (\ref{WKB1}) and (\ref{WKB2}).
From (\ref{WKB1}) we see that $S_{-1}(x)$ has a pole of order 1 at $x=1$, while from (\ref{WKB2}) we see that $S_j(x)$
with $j\ge 0$ has a pole at $x=1$.
Consequently, for all $j$ we have the Laurent expansions
\bea
   S_j(x)={f_{j,-1}\over x-1}+\sum_{n\ge 0}f_{j,n} (x-1)^n\label{laurent} 
\eea
Thus, the residue of $S(x,N)$ at $x=1$ is given by
\bea
  {\rm Res}_{x=1}\, S(x,N)=\sum_{j=-1}^\infty f_{j,-1} N^{-j}
\eea
To perform the sum on the right hand side, plug (\ref{laurent}) into (\ref{powerser}), and then plug (\ref{powerser}) into the Riccati equation. Equating the coefficient of $(x-1)^{-2}$ to zero, we find
\bea
   -\sum_{j=-1}^\infty f_{j,-1} N^{-j}+\left( \sum_{j=-1}^\infty f_{j,-1} N^{-j}\right)^2=N^2\rho(N)^2
\eea
This quadratic equation is easily solved to obtain
\bea
   \sum_{j=-1}^\infty f_{j,-1} N^{-j}={1\over 2}+N\rho(N) \sqrt{1+{1\over 4 N^2\rho(N)^2}}={\rm Res}_{x=1}\, S(x,N)
\eea
For the residue of $S_{\rm odd}$ we need to extract the odd powers of $N$ which gives
\bea
{\rm Res}_{x=1}\, S_{\rm odd}(x,N)={1\over 2}\left(N\rho(N) \sqrt{1+{1\over 4 N^2\rho(N)^2}}
+N\rho(-N) \sqrt{1+{1\over 4 N^2\rho(-N)^2}}\right)
\eea
For even $\rho(N)$ we recover the result of \cite{KawaiTakei} which says
\bea
{\rm Res}_{x=1}\, S_{\rm odd}(x,N)=N\rho(N) \sqrt{1+{1\over 4 N^2\rho(N)^2}}
\eea

\end{document}